# Undulations of smectic A layers in achiral liquid crystals manifested as stripe textures

Natalia Podoliak,[1] Peter Salamon,[2] Lubor Lejček,[1] Petr Kužel[1] and Vladimíra Novotná[1]

[1] *Institute of Physics of the Czech Academy of Sciences, Na Slovance 2, CZ-182 00 Prague 8, Czech Republic*
[2] *Institute for Solid State Physics and Optics, Wigner Research Centre for Physics, P.O. Box 49, Budapest H-1525, Budapest, Hungary*

Self-assembly of organic molecules represents a fascinating playground to create various liquid crystalline (LC) nanostructures. In this work, we study layer undulations on micrometer scale in smectic A phases for achiral compounds, experimentally demonstrated as regular stripe patterns induced by thermal treatment. Undulations including their anharmonic properties are evaluated by means of polarimetric imaging and light diffraction experiments in cells with various thicknesses. The key role in stripe formation is played by high negative values of the thermal expansion coefficient.

Diverse number of physical and biological systems can form microscopic patterns on macroscopic scale, spontaneously or under the influence of external stimuli. Stripe patterns could be observed in modulated phases in biological membranes [1-3], thin films of block polymers [4], lamellar liquid crystals (LCs), which are of utmost interest to us. Indeed, thermotropic LCs are an example of soft matter, which reveal intermediate phases (mesophases) between the isotropic liquid and the solid crystal in a certain temperature range [5]. Due to the anisotropy of the physical properties, LCs show various enthralling textures under a polarizing microscope, depending on the surface alignment and the orientation of molecules in the cell, which offers a possibility of numerous structural modifications. LCs are very sensitive to external stimuli such as stress, electric or magnetic field, which enables controlling the functional properties in most applications.

Smectic LC phases are layered structures with molecules oriented along the layer normal, smectic A (SmA) phase, or tilted with respect to it, smectic C (SmC) phase. While in nematic LCs the generation of regular modulated structures can easily be done e.g. by electric field [6-9], the emergence of stripe patterns is to a great extent less common in the case of smectics.

When layered systems in confined geometries are subjected to dilatation perpendicular to the layers, the layer modulation takes place, manifesting itself by stripes in microscope images. Stripe textures have been observed in SmA phase under external stress in the geometry with the layers parallel to the cell surfaces (homeotropic geometry) [5,10-12], in confined geometries with curved [13-17] or flat [18] boundaries. Besides, stripe textures have been observed in chiral electroclinic LCs in planar geometry (with smectic layers perpendicular to the surfaces) near the phase transition to the SmC* phase under an external electric field [19-21].

The problem of undulations has been also treated in hexagonal columnar mesophases of lyotropic systems [10,22,23]. In the SmA phase of achiral molecules, a periodic pattern induced by an applied electric field was reported in a special confinement between electrodes under conflicting anchoring conditions [24]. However, without external stimuli, a regular stripe pattern has not been reported yet in the SmA phase. An unusual optical effect was observed in the SmA phase in Ref. [25] for a very thick sample. The observed scattered-light crescents at different incident angles were explained by defects at the surfaces.

For chiral rod-like molecules or achiral bent-core mesogens, modulated structures spontaneously appearing without external stimuli can be observed, being caused by molecular or structural chirality. A twist-bend nematic (NTB) phase was discovered for achiral dimers [26]. For the NTB phase, a heliconical structure with a periodicity of about 10 nm was proved. The thermal evolution of this pseudo-layer structure causes the appearance of stripes on a micrometer scale, and the stripe periodicity was found to be twice the cell thickness [27,28]. For similar kind of achiral dimeric molecules, a regular stripe pattern was found also in the SmC phase [29].

A periodic variation of the optical axis orientation in a modulated sample constitutes a polarization-sensitive diffraction grating. Spontaneously formed polarization gratings were reported and analyzed in the NTB phase [30,31], a tunable optical grating made of flexoelectric domains was observed in a bent-core nematic liquid crystal [32], and a two-dimensional diffraction grating was reported in the SmC phase [33].

All the above-described modulated structures were subjected to intensive theoretical efforts. Various models and approaches were proposed to explain the observed modulations: from small sinusoidal undulations [10,20,34] to zig-zag [21,35,36] and soliton-like structures [19]; even more complex problem of transition of sinusoidal undulations to chevron-like buckling (non-linear) undulation profile was discussed by Singer et al. [22].

In the present work, we report on the experimental evidence of the regular stripe pattern in homogeneously aligned cells (in planar geometry) of achiral compounds and racemic mixtures of chiral LC compounds in the SmA phase. The stripe pattern, observed under polarizing microscope, appears upon the sample heating and is described by the smectic layer modulations enabled by a rather high negative value of the thermal expansion coefficient. The polarimetric imaging and polarized light diffraction experiments allow us to estimate the amplitude, period and higher-harmonic components of the undulations. A theoretical model is proposed to describe the observed behavior.

Herein, we analyze the effect for achiral LC denoted as II/6 and for a racemic mixture of enantiomers denoted as 9ZBL. The chemical formulas, phase transition temperatures and the phase sequences of the studied materials are summarized in Supplemental Material (SM). The material synthesis and their mesogenic properties have been described in Refs. [37] for II/6 and [38,39] for 9ZBL. Both materials exhibit the SmA phase in a broad temperature range (more than 50 K). Above the studied SmA phase, the presence of a nematic (N) phase is essential to achieve a good alignment of molecules and layers.

We utilized commercial cells of several thicknesses ($d$ = 1.7, 3.0, 5.0, 8.0 and 11.6 μm), purchased from WAT PPW company, with glasses covered with transparent ITO electrodes and a surfactant, which ensured uniform planar alignment. The cells were filled with the studied compounds in the isotropic phase (Iso) by means of capillary action. After cooling from Iso to N and then to the SmA phases, perfectly homogeneously aligned planar textures were obtained as it was observed under the microscope (see photos in SM). After cooling the cell deeply into the SmA phase, we stopped cooling and applied small heating of the sample (of only 0.1 K as measured on a heating stage). After that, regular stripes appeared in the whole sample area parallel to the rubbing direction (see photos in SM), as it is schematically shown in Fig. 1. The contrast of the stripes increased on further heating.

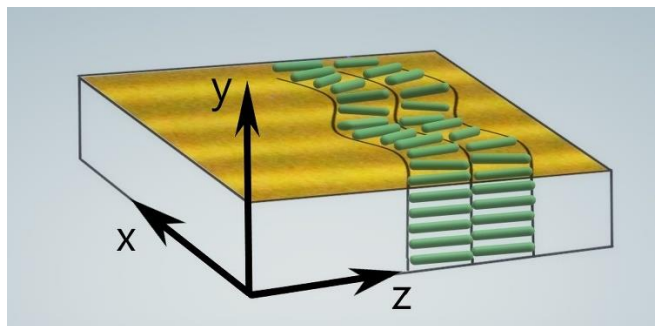

FIG. 1. Geometry of the studied system in the SmA phase in the sample frame: $z$-axis is perpendicular to the smectic layers and is along the rubbing direction of the cell; layer undulations along x-axis appear due to the temperature gradient along y-axis.

The periodic pattern of the stripes does not change when the temperature is stabilized and upon further heating or cooling of the sample. The stripe textures disappeared only when leaving the SmA phase on cooling or heating. As the rubbing direction of the cell corresponds to the long molecular axis, the smectic planes are perpendicular to the stripe orientation. Almost all experimental results presented here are from 9ZBL; however, similar results were found also for II/6.

We studied the stripe textures of LC compounds with respect to the cell thickness. For a thin cell ($d = 1.7$ μm), a periodic pattern with a periodicity of about 1.5-2 μm was observed (Fig. 2(a)). In general, for cells thinner than 5 μm, the stripe textures did not significantly change upon further heating or cooling. For thicker cells, the situation was more complicated, as it is demonstrated in Fig. 2(b) for a 5-μm thick cell (patterns for cells with other thicknesses are shown in Fig. S4 in SM). For these samples, focal conics were growing upon further heating and for the cell thickness of 11.6 μm they dominated in the textures (see SM).

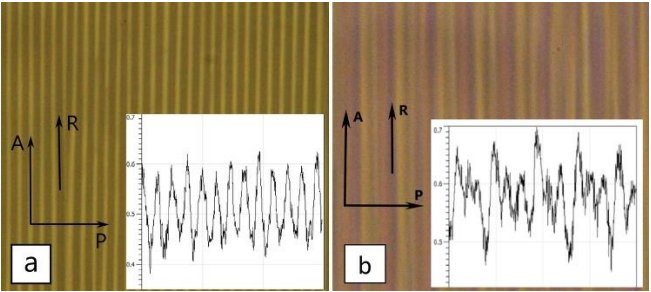

FIG. 2. Planar textures of 9ZBL in the SmA phase between crossed polarizers (shown by arrows A and P) in: (a) 1.7 μm cell; (b) 5.0 μm cell. The long molecular axis is directed along R (rubbing direction). The width of the photos is about 40 μm. The intensity profiles in the direction perpendicular to the stripes are shown in the insets.

The stripe patterns represent layer undulations with the wavelength $\Lambda$ being twice the stripe periodicity. We analyzed the intensity profiles of the polarized microscope photos in the direction perpendicular to the stripes and performed the Fourier analysis of the patterns for 9ZBL samples of several thicknesses (see Fig. S5 in SM). We note small inhomogeneities in the sample properties, which were recognized from the spatial dependence of the undulation period in the thinnest sample. Such inhomogeneities were not observed in thicker samples; instead, higher harmonics are well developed in the Fourier spectra, which demonstrate certain anharmonicity of the undulation profile. The compilation of wavelengths $\Lambda$ obtained from the Fourier analysis is plotted in Fig. 3.

In order to assess the possible influence of the surface treatment, we checked our results by using various commercial cells from other companies (Linkam, etc.): similar stripe patterns were always observed upon heating the samples in SmA phase.

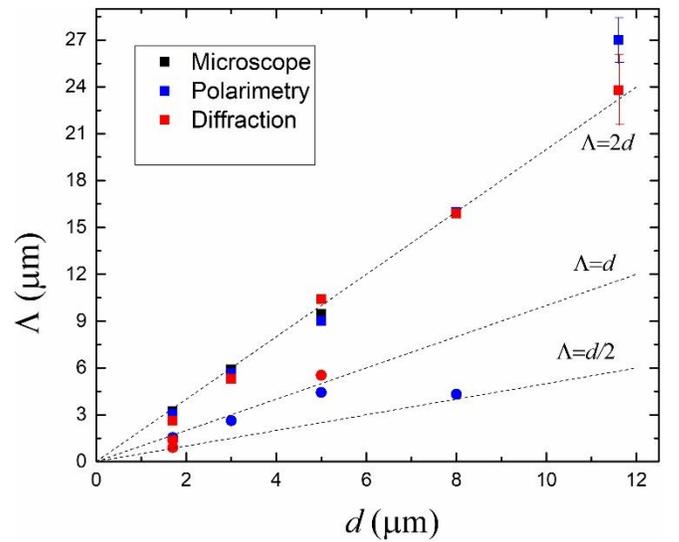

FIG. 3. The wavelength of the undulations, $\Lambda$, as a function of the sample thickness, $d$, obtained using several experimental techniques: polarized microscope images (black), polarimetric data (blue) and light diffraction experiments (red). Fundamental wavelengths are shown with squares and certain higher harmonics with circles.

We studied the distribution of local birefringence by means of imaging polarimetric measurements in monochromatic light with $\lambda = 657.7$ nm. We used a setup allowing us to prepare the input beam with the well-defined elliptical polarization state and analyze its changes upon propagation through the sample with the spatial in-plane resolution yielding 64×64 pixels images on a scale of tens of micrometers. The pixelwise maps of the net phase retardation $\Delta\Phi$ and of the azimuthal angle $\varphi$ defining the in-plane slow axis direction are shown in $x'z'$ coordinates. The laboratory frame $x', y, z'$ is rotated by 45° around $y$ axis with respect to the sample frame introduced in Fig. 1. For details about the experimental method, see [40] and SM.

An example of the distribution of the azimuthal angle $\varphi$ is shown in Fig. 4(a) for a 3.0 μm cell of 9ZBL. The colored scale corresponds to the projection of the director (averaged along $y$) onto the sample plane and yields the variation of the director component in the sample plane. The maximum in-plane molecular tilt from the mean director orientation inferred from this experiment is 4 deg. A more complicated stripe structure with additional harmonics is clearly seen for a 5 μm sample, see Fig. 4(b). The polarimetric results for all the samples were analyzed by Fourier transformation to obtain the periods of the modulations (see SM). The results of this analysis are summarized in Fig. 3.

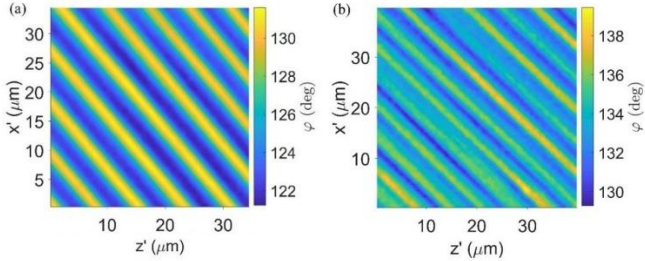

FIG. 4. Color map of the azimuthal angle $\varphi$ (a) for a 3 μm cell and (b) for a 5 μm cell.

The studied stripe array acts as a diffraction grating for linearly polarized light. The geometry of the diffraction experiment and the theoretical background are described in SM. An example of the diffraction pattern observed for the 1.7 μm 9ZBL sample is shown in Fig. 5. The polarization of the directly transmitted beam (0[th] order) was parallel to that of the incident beam (Fig. 5 shows only intensity decrease in crossed polarizer geometry); the diffracted beams (observed up to the 4[th] order; 3[rd] one is not visible) showed a significant linearly polarized component in the perpendicular direction. Higher order diffraction spots reflected the anharmonicity of the undulations. The Fourier analysis of the polarimetric measurements for the 1.7 μm 9ZBL sample confirmed the weakness of the third harmonic component in the undulation pattern, see SM. For thicker samples, the diffraction patterns are even more complex and certain higher order peaks are more intense than the first order peak. The diffraction pattern for the 5 μm 9ZBL sample is shown in SM.

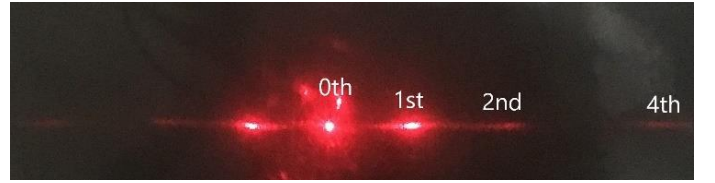

FIG. 5. Diffraction pattern for 9ZBL compound between crossed polarizers for a 1.7 μm sample.

From the measured intensities and positions of the diffraction spots on the screen, we determined the parameters of the undulated layers (see details in SM) for selected cell thicknesses. The calculated wavelengths, Λ, of the undulations, and their amplitudes, $A$, are collected in Table I for certain cell thicknesses. The values of Λ are plotted as a function of the cell thickness in Fig. 3 (red points).

The diffraction data in Table I indicate that the undulation amplitudes are on a sub-100-nm scale and they decrease with increasing of the sample thickness, which means that in "squeezed" conditions of thin cells the layer undulations are more pronounced. By simple geometrical calculations, the maximal deviation of the molecules from the alignment in the modulated system was estimated to be 3-4 deg, which is in a perfect agreement with the variation of $\varphi$ shown in Fig. 4(b).

As described above, we used three independent methods to determine the periodicities of the undulations in the samples and the results are summarized in Fig.3. Apparently, the fundamental undulation periods depend linearly on the cell thickness as $\Lambda \cong 2d$. Besides, higher harmonics are also visible, being sometimes even more intense in thicker samples than the fundamental ones.

TABLE I. The wavelengths of the undulations (Λ), and their amplitudes ($A$), obtained from the diffraction experiments for selected cell thicknesses of 9ZBL compound.

| $d$ (μm) | Λ (μm) | $A$ (nm) |
|---|---|---|
| 1.7 | 2.7±0.8 | 80±20 |
| 5.0 | 10.5±1.5 | 65±15 |
| 11.6 | 23.5±2.0 | 40±10 |

The key driving force of the studied effect is related to the thermal expansion. We calculated the thermal expansion coefficients based on the temperature dependence of the layer spacing (obtained from x-ray measurements [37-39]): $k_t = -0.8 \times 10^{-3}$ nm/K for II/6 and $k_t = -1.8 \times 10^{-3}$ nm/K for 9ZBL. Indeed, it is typical that the layer thickness linearly grows on cooling (negative value of $k_t$) in SmA phases; this is related to the terminal chains stretching and/or to an increase in the orientational order of the molecules.

To explain the stripe appearance in the studied system, we propose a model considering small sinusoidal undulations similarly as in the previously studied systems [5,10,22,23,34]. During the experiments, the heater was in contact with the lower surface of the cell and a temperature gradient inside the sample could develop during the temperature changes. We denote by $T_d$ and $T_o$ the temperatures at the upper and lower sample surfaces, respectively. During the sample cooling, the layer thickness gradually increases with decreasing temperature below the phase transition N-SmA. When the sample cooling is stopped, the sample temperature is stabilized at the value $T_d$. During the subsequent sample heating, the temperature of the bottom part of the cell is higher than the upper part $(T_o - T_d) > 0$ and the thickness of the smectic layers near the bottom surface becomes smaller than that at the top one, $a_0 \equiv a(T_o) < a(T_d) \equiv a_d$, due to the negative thermal expansion coefficient $k_t$. As a result of such dilatation, the layer undulations might start to develop at the bottom surface.

The elastic free energy density per unit volume, $f$, including the second-order anharmonic correction for the modulated state was proposed previously in Ref. [10] in the following form:

$$f = \frac{B}{2}\left(\frac{\partial u}{\partial z} - \frac{1}{2}\left(\frac{\partial u}{\partial x}\right)^2 - \frac{1}{2}\left(\frac{\partial u}{\partial y}\right)^2\right)^2 + \frac{K}{2}\left(\frac{\partial^2 u}{\partial x^2} + \frac{\partial^2 u}{\partial y^2}\right)^2. \quad (1)$$

The layer displacement along the $z$-axis is denoted as $u(x, y, z)$, $B$ is the layer modulus of compressibility and $K$ is the curvature modulus. A characteristic elasticity length of the material, $\lambda_{\text{el}}$, was introduced by de Gennes [5] as $\lambda_{\text{el}} = \sqrt{K/B}$. The layer displacement $u(x, y, z)$ can be written in the form:

$$u(x, y, z) = \gamma(y)z + A\cos(k_x x)\sin(k_y y), \quad (2)$$

where $k_x = 2\pi/\Lambda$ is the wave-vector of undulations along $x$-axis (with the undulation wavelength $\Lambda$) and $k_y$ is connected to the sample thickness $d$: $k_y = \pi/d$. The parameter $A$ is the undulation amplitude. Note that the function $\sin(k_y y)$ satisfies the fixed boundary conditions at $y = 0$ and $y = d$. Since $k_t < 0$, the layer thickness decreases with increasing temperature. The dilatation $\gamma = (a(T_d) - a(T))/a(T)$ is temperature dependent and varies along $y$-axis $\gamma(y) = \gamma_o y/d$, where $\gamma_o$ is the dilatation at $T_o$. We denote by $l_z$ a typical correlation length of undulations in the direction normal to the smectic layers. The dilatation gradient would cause a total displacement $\gamma(d)l_z$ of smectic molecules for several tens of smectic layers, which is energetically costly. Therefore, the undulations described by the second term in Eq. (2) with similar displacement magnitude (Table 1) may stabilize the structure. A smectic segment of the size $l_z$ along $z$-direction can be periodically repeated, which permits the evaluation of the average free energy in an analogy to Born-von Karman periodic boundary condition. Introducing (2) into (1), one can calculate the average free energy per unit area of the sample:

$$<f> = \frac{1}{\Lambda l_z} \int_0^\Lambda dx \int_0^d dy \int_0^{l_z} f dz. \quad (3)$$

The undulations can exist if $<f>$ exhibits a minimum. We found (see SM for details) that this is fulfilled if the dilatation exceeds a critical value, $\gamma_o \geq \gamma_{oc}$:

$$\gamma_{oc} = 5\pi^2 \frac{5-4q}{q^2}\left(\frac{\lambda_{\text{el}}}{d}\right)^2, \quad (4)$$

where $q$ is an adjustable bound parameter ($0 < q < 1$) connected to $l_z$:

$$l_z = q\frac{\sqrt{3}d^2}{5\pi\lambda_{\text{el}}}. \quad (5)$$

The undulation wavelength is then expressed as:

$$\Lambda = \frac{2d}{\sqrt{5}\sqrt{1/q - 1}}. \quad (6)$$

Since the parameter $q$ does not depend on the sample thickness $d$ ($l_z$ does), the period $\Lambda$ is linearly proportional to $d$ as inferred from (6). In particular, for $\sqrt{5(1/q - 1)} = 1$, which provides $q = 5/6$, one obtains exactly $\Lambda = 2d$. This is a non-trivial assumption; however, it gives us a good correspondence with the experimental results.

The thicker the sample, the more spatial harmonics were observed in undulations. In some cases, a particular higher harmonic was found to be more intense than the fundamental component. Undulations involving higher harmonics in SmA phase were previously reported only in a chiral system under applied electric field [41]. Our harmonic model (2) cannot provide a quantitative assessment and its generalization would bring a large number of unknown parameters. Besides that, more complicated structures may also appear and play a significant role in the case of thick samples.

With further heating, the layer deformations increase and undulations might change into buckling deformations [22]. The layer buckling becomes more and more pronounced, which leads to the creation of parabolic focal conic domains appearing within the system of stripes [10]. With the temperature increase, their number increases and the stripe texture changes completely into the focal conic domain texture, as it is demonstrated in Fig. S4(h).

To conclude, we have found a regular pattern of stripes for achiral liquid crystalline compounds in the SmA phase. To our best knowledge, it is the first system with undulated structure in the SmA phase, which is induced only by external thermal influence. As the stripes are found in the materials with a large negative value of the thermal expansion coefficient, $k_t$, we assume that the material is subjected to frustration between the upper and bottom parts of the cell during heating. Due to the viscosity and the molecular anchoring at the bottom surface, the layer contraction is large enough to induce the periodic undulated structure. To explain the observed stripe patterns, we developed a model based on the critical deformation for undulations. The stripe system is very stable and remains even on subsequent cooling or heating. The modulation period was established to be proportional to the cell thickness, which can be useful for the variability of diffraction gratings in photonic devices.

The proposed explanation of modulations can serve as a model system for thermally induced expansion and changes in various polymeric or biologically important layered structure.

Authors acknowledge the project FerroFluid, EIG Concert Japan - 9[th] call "Design of Materials with Atomic Precision". S.P. acknowledges the project NKFIH FK142643 and the support of the János Bolyai Research Scholarship of the Hungarian Academy of Sciences.